\documentclass[conference]{IEEEtran}
\makeatletter
\newcommand{\newlineauthors}{%
  \end{@IEEEauthorhalign}\hfill\mbox{}\par
  \mbox{}\hfill\begin{@IEEEauthorhalign}
}
\makeatother
\IEEEoverridecommandlockouts
\usepackage{cite}
\usepackage{amsmath,amssymb,amsfonts}
\usepackage{algorithmic}
\usepackage{graphicx}
\usepackage{textcomp}
\usepackage{xcolor}
\usepackage{wrapfig}
\usepackage{algorithmic}
\usepackage{graphicx}
\usepackage{textcomp}
\usepackage{xcolor}
\usepackage{array}
\usepackage{tabularx}
\usepackage{ctable}
\usepackage{threeparttable}
\usepackage{color, colortbl}
\usepackage{caption,subcaption}
\usepackage{balance}

\definecolor{Gray}{gray}{0.90}

\def\BibTeX{{\rm B\kern-.05em{\sc i\kern-.025em b}\kern-.08em
    T\kern-.1667em\lower.7ex\hbox{E}\kern-.125emX}}
\begin{document}

\title{Workload Failure Prediction for Data Centers}

\author{\IEEEauthorblockN{Jie Li}
\IEEEauthorblockA{\textit{Department of Computer Science} \\
\textit{Texas Tech University}\\
Lubbock, USA \\
jie.li@ttu.edu}
\and
\IEEEauthorblockN{Rui Wang}
\IEEEauthorblockA{\textit{Department of Computer Science} \\
\textit{Texas Tech University}\\
Lubbock, USA \\
rui.wang@ttuhsc.edu}
\and
\IEEEauthorblockN{Ghazanfar Ali}
\IEEEauthorblockA{\textit{Department of Computer Science} \\
\textit{Texas Tech University}\\
Lubbock, USA \\
Ghazanfar.Ali@ttu.edu}
\newlineauthors
\IEEEauthorblockN{Tommy Dang}
\IEEEauthorblockA{\textit{Department of Computer Science} \\
\textit{Texas Tech University}\\
Lubbock, USA \\
tommy.dang@ttu.edu}
\and
\IEEEauthorblockN{Alan Sill}
\IEEEauthorblockA{\textit{High-Performance Computing Center} \\
\textit{Texas Tech University}\\
Lubbock, USA \\
alan.sill@ttu.edu}
\and
\IEEEauthorblockN{Yong Chen}
\IEEEauthorblockA{\textit{Department of Computer Science} \\
\textit{Texas Tech University}\\
Lubbock, USA \\
yong.chen@ttu.edu}
}

\maketitle

\begin{abstract}
  Failed workloads that consumed significant computational resources in time and space affect the efficiency of 
  data centers significantly and thus limit the amount of scientific work that can be achieved. While the computational power has increased significantly over the years, 
  detection and prediction of workload failures have lagged far behind and will become increasingly critical as the system scale and complexity further increase.
  In this study, we analyze workload traces collected from a production cluster and train machine learning models on a large amount of data sets to predict workload failures. Our prediction models consist of a queue-time model that estimates the probability of workload failures before execution and a runtime model that predicts failures at runtime. Evaluation results show that the queue-time model and runtime model can predict workload failures with a maximum precision score of 90.61\% and 97.75\%, respectively. By integrating the runtime model with the job scheduler, it helps reduce CPU time, memory usage by up to 16.7\% and 14.53\%, respectively.
\end{abstract}


\section{Introduction}

The scale and complexity of many data centers have significantly increased over the years. Meantime, the demand from user community for computational and storage capability has considerably increased too. This combination of increased scale of 
data centers and size of workloads with different requirements and characteristics has resulted in growing node and workload failures, posing a threat to the reliability, availability, and scalability (RAS) of data centers. 
For example, all else being equal, a system that is 1,000 times more powerful will have at least 1,000 times more components and will fail 1,000 times more often~\cite{cappello2014toward}, resulting in a long-running job utilizing a large amount of nodes being terminated due to frequent failures. 
Therefore, over the past decades, various methods and algorithms were proposed to improve the system resilience and efficiency~\cite{candea2004recovery, candea2004microreboot, hargrove2006berkeley, garg2011environment, aupy2016co, rodriguez2019job}. 

Reactive strategies, such as Checkpoint/Restart (C/R)~\cite{hargrove2006berkeley, rodriguez2019job}, are conventional approaches for fault tolerance. As an example, a reactive fault tolerance strategy for a node failure is to reschedule a workload to a new node and restart from a specific checkpoint. However, checkpointing a job in a large-scale system could incur large I/O overhead when writing and reading workloads state~\cite{garg2018shiraz}, and takes an overhead of more than 15\% of the total execution time~\cite{elnozahy2004checkpointing, cappello2009fault}, which significantly impedes science productivity. As a result, researchers on failure management have found that prevention is better than cure and shifted to proactive management strategies~\cite{sahoo2003critical, yalagandula2004beyond, mickens2006exploiting, nukada2011nvcr, rezaei2014snapify, rodriguez2019job}. In contrast to reactive strategies, proactive strategies develop models based on the failure data in data centers 
to predict node or workload failures in the near future and take preventive measures to improve the RAS of data centers. 

Numerous research efforts have developed node failure detection and prediction methods by utilizing temporal and/or spatial correlations of failures~\cite{ el2013reading, ghiasvand2016lessons, kimura2018proactive, ghiasvand2019anomaly}. They usually investigate system behavior via Syslog analysis and have developed supervised and unsupervised approaches for predicting failures in data centers. 
A number of studies have attempted on workload-centric failure detection and prediction based on the resource usage or requested resources~\cite{fadishei2009job, chen2014failure, islam2017predicting, andresen2018machine}. However, only limited amount of workload data is publicly available due to confidentiality or other reasons. In addition, analyzing and extracting insightful knowledge from massive amounts of data is daunting, given the increasing scale and complexity of data sets.

This research aims at using machine learning-based approach to predict workload failures in data centers. 
In particular, we investigate two months of workload traces collected from a production cluster in order to find correlation between workload attributes with exit status (including error status). We seek to train supervised learning models to predict: (1) the failure probability of a workload at queue time, and (2) the likelihood of failure over the life-span of a workload. Having the knowledge of whether jobs will likely to fail or not can be valuable for both users to be alerted of the potential failures and the resource manager (both the software and system administrators) to be proactive in preventing wasting computational resources. Consequently, the RAS and productivity of data centers can be improved in return by better managing the workloads that are likely to fail. 

We make the following contributions in this study:
\begin{itemize}
    \item We analyze workload traces collected from a production data center and perform an extensive characterization study of workload failure rates across nodes, users and different time scales. We investigate the correlation between workload characteristics and failures, and identify the relevant factors that lead to failures.
    \item We apply several machine learning algorithms on our data set and train two prediction models: a \emph{Queue-time model} and a \emph{Runtime model}. We find that \emph{Random Forest} achieved the best prediction performance in terms of Precision and F1 scores in both models. Experimental results show that these two models predict workload failures with a maximum Precision of 90.61\% and 97.75\%, respectively.
    \item We quantify the resource savings achieved by applying the runtime prediction model on workloads at different times. Based on the prediction results, proactive failure management (e.g., killing workloads that are predicted to fail) can achieve CPU and memory savings by up to 16.7\% and 14.53\%, respectively.
    \item We investigate the effects of training data set size to find the optimum size that can achieve acceptable prediction performance with minimum training time.
\end{itemize}

The rest of this paper is organized as follows. 
Section~\ref{priliminaries} describes background of this research, including the monitoring infrastructure, data points, and source of anomalies. In Section~\ref{analysis}, we analyze the workload data. Section~\ref{models} describes the machine learning algorithms we have investigated in this research and explains our methodology. The experimental results are presented in Section~\ref{results}. Section~\ref{related} provides an overview of related work, and we conclude this research in Section~\ref{conclusion}.

\section{Background}\label{priliminaries}

This research study is conducted on a production data center called \emph{Quanah}, where scientists from all major scientific fields, such as astrophysics, computational chemistry, bioinformatics, etc., perform simulations and scientific computations. In our previous work~\cite{li2020monster} , we have designed and implemented a monitoring, data collection and management infrastructure to gather workload and node metrics from the cluster in real time. The Quanah cluster and monitoring framework are described in the next section, followed by the analysis of sources of failures in common data centers. 

\subsection{Quanah Cluster}

The Quanah cluster at High Performance Computing Center (HPCC) of Texas Tech University~\cite{hpcc}
is commissioned in early 2017 and expanded to its current size in 2019, which is comprised of 467 nodes with Intel XEON processors providing 36 cores per node. Quanah has a total of 16,812 cores with a benchmarked total computing power of 485 Teraflops/s and provides 2.5 petabytes storage capability. The cluster is based on Dell EMC PowerEdge™ C6320 servers, which are equipped with the integrated Dell Remote Access Controller (iDRAC)~\cite{idrac} providing Redfish API~\cite{redfish} for accessing Baseboard Management Controller (BMC). The software environment is based on CentOS 7 Linux, provisioned and managed by OpenHPC, and the cluster is operated with Univa Grid Engine (UGE)~\cite{uge}, setting up with multiple queues, with jobs sorted by projects to meet the needs of research activities for many fields and disciplines. 

\subsection{Monitoring Infrastructure}\label{monitoring}

The monitoring data in Quanah cluster is obtained through an ``out-of-the-box'' monitoring tool~\cite{li2020monster}
that utilizes the Redfish API to retrieve sensor data from the BMC on each compute node and the resource manager (such as UGE and Slurm) for workload information and resource usage data. Sensor metrics and resource usage data are collected at node level in 60-second intervals, which include power usage, fan speed, CPU usage, memory usage and node-job correlations, etc. The time-series data is stored in a time-series database (e.g. InfluxDB). Workload information is derived from the UGE accounting data, which includes job submission time, start time, end time, total CPU time, integral memory usage, IO operations, etc. The workload data is stored in a MySQL database. With several performance optimizations, such as optimized database schema, using high-speed storage, concurrent processing and transmitting compressed data, our infrastructure provides near real-time analysis and visualization of user-level and node-level status.  

\subsection{Sources of Failures}
\begin{figure}[t]
\centering
\includegraphics[width=0.95\linewidth]{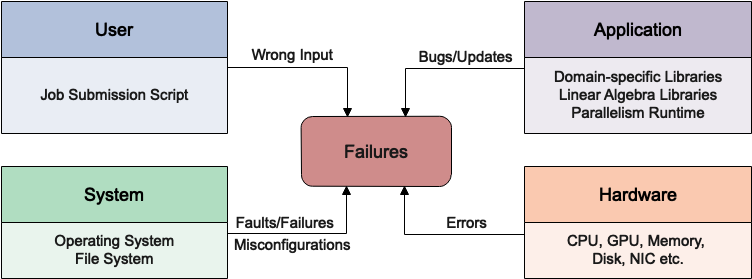}
\caption{Sources of Failures in Data Centers}
\label{fig:sources}
\end{figure}

In a data center cluster where computing resources are shared by different workloads submitted by users from various domains, the number of failed jobs (i.e. workloads) can be large. There are three main reasons. First, domain scientists, while skilled in their scientific fields, do not always have sufficient experience and background in computing, especially in large-scale, parallel computing. Second, diverse workloads depend on different libraries, and bugs and missing updates in dependent libraries can lead to unexpected failures. Third, data center is complex, and any mis-configuration or hardware errors can cause workload termination. 

Figure~\ref{fig:sources} summarizes the high-level root cause categories in data centers, where failures are attributed to user, application, system or hardware problems. 
\begin{itemize}
    \item \emph{Users}: insufficient resource request or wrong input in the job submission script can cause workloads to fail~\cite{li2006job}. Note that user interruptions, such as issuing a command like ``scancel'', can interrupt a running workload and cause a failure too. However, since the effect of cancelling a running workload is obvious to the user, we do not categorize it as a source of failures.
    
    \item \emph{Applications}: mis-configured applications increase the risk of poor performance. In addition, buggy codes, missing dependent library updates, and/or bugs can cause applications to terminate unexpectedly too. 
    
    \item \emph{Systems}: mis-configurations or failures of system resources and components may considerably affect the performance of workloads or cause failures. 
    
    \item \emph{Hardware}: hardware errors are one of the most devastating issues for data centers. Severe hardware issues can lead to the malfunction of the entire system. Events such as memory hardware errors, CPU overheating, etc., will result in workload errors and crashes. 
\end{itemize}

\section{Workload Analysis}\label{analysis}

To predict the workload failures in data centers, it is crucial to understand the characteristics of failed workloads. In this section, we first present an overview of the workload trace, then quantitatively analyze the percentage of failures in the workloads and the computational resource consumption characteristics. After that, we further study the failure rate across the nodes, users, and different time scales.

\subsection{Workload Overview}\label{workload}

\begin{table}
    \centering
    \caption{Features of Workloads}
    \begin{threeparttable}
    \begin{tabular}{r c l}
    \specialrule{.1em}{.05em}{.05em}
        \rowcolor{Gray}
         \textbf{Feature} & \textbf{Type} &\textbf{Description}  \\
    \specialrule{.1em}{.05em}{.05em}
         job\_id & Numeric & Job identifier \\
         owner & Categorical & Owner of the job \\
         group & Categorical & The group id of the job owner \\
         job\_name & Categorical & Job name \\
         granted\_pe & Categorical & The parallel environment \\
         hostname & Categorical & Name of the execution host \\
         submission & Numeric & Submission time (in epoch time format)\\
         start\_time & Numeric & Start time (in epoch time format)\\
         end\_time & Numeric & End time (in epoch time format) \\
         wallclock & Numeric & Difference between end and start time \\
         cpu & Numeric & The CPU time usage in seconds \\
         mem & Numeric & The integral memory usage in Gbytes\tnote{1}\\
         io & Numeric & The amount of data transferred in Gbytes \\
         iow & Numeric & The io wait time in seconds \\
         maxvmem & Numeric & The maximum vmem size in bytes \\
         slots & Numeric & The number of parallel processes \\
         wait\_time & Numeric & Difference between start and submit time \\
         exit\_status & Numeric & Exit status of the job script \\
    \hline
    \end{tabular}
    \begin{tablenotes}
      \small
      \item[1] The sum of the amount of memory used in each time interval for the life of the job.
    \end{tablenotes}
    \end{threeparttable}
    \label{tab:features}
\end{table}

\begin{table}
    \centering
    \caption{Exit Status Summary for Failed Workloads}
    \label{tab:exit}
    \begin{tabular}{c|c|c|c}
        \specialrule{.1em}{.05em}{.05em}
        \rowcolor{Gray}
        \textbf{Exit Code} & \textbf{Meaning} & \textbf{Number} & \textbf{Percentage}\\
        \specialrule{.1em}{.05em}{.05em}
        1 & Miscellaneous errors & 21367 & 58.16\% \\
        \hline
        2 & Missing keyword or command & 3032 & 8.25\% \\
        \hline
        7 & Argument list too long & 6549 & 17.83\% \\
        \hline
        127 & Command not found & 528 & 1.44\% \\
        \hline
        137 & (System signal 9) Kill & 190 & 0.52\% \\
        \hline
        255 & Exit status out of range & 4598 & 12.52\% \\
        \hline
        \multicolumn{2}{c|}{Others} & 475 & 1.28\% \\
        \hline
    \end{tabular}
\end{table}

\begin{figure}
    \centering
    \includegraphics[width=0.95\linewidth]{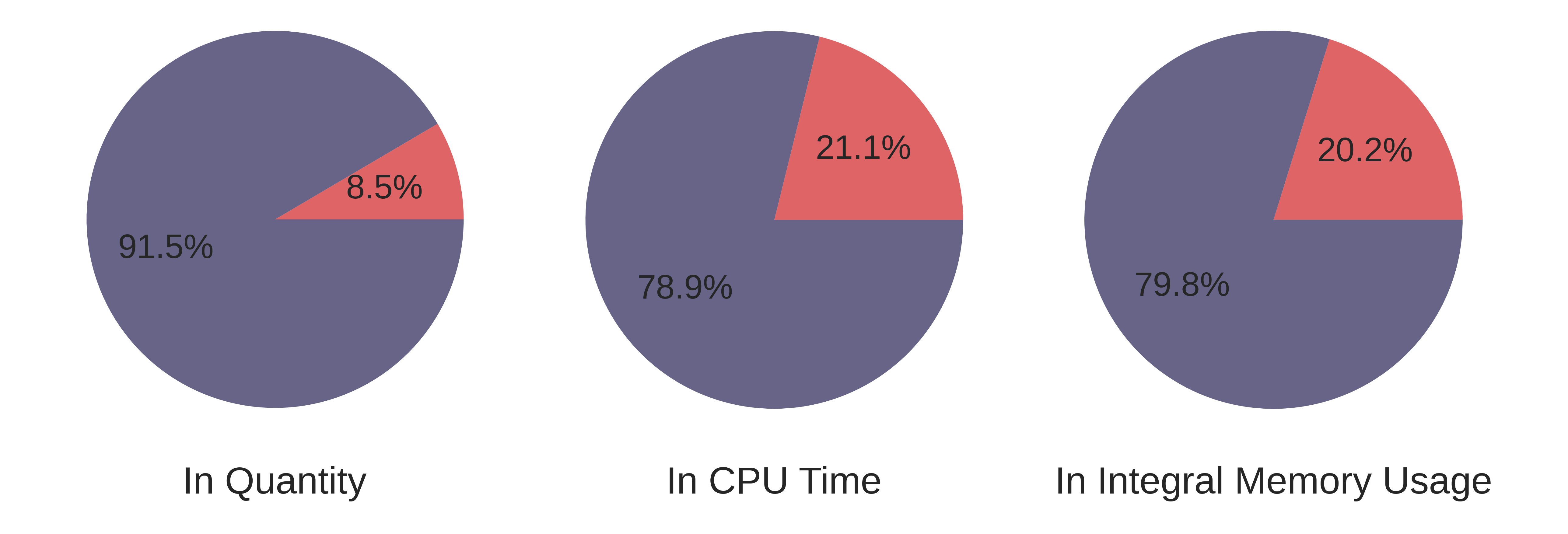}
    \caption{Proportion and Resource Consumption Characteristics of workloads. Red indicates failed workloads and dark blue indicates successful and cancelled workloads (non-failures).}
    \label{fig:abnormal}
\end{figure}

The workload trace is derived from job accounting data collected from the Quanah cluster for the period of August 1, 2020 to October 1, 2020, which contains 324,358 instances submitted by 204 unique users (i.e. owners). Notice that workloads that cannot be started on the execution host (e.g., because the user does not have a valid account on that node~\cite{ugeaccounting}) are recorded in the raw job accounting data, and we drop these entries because they are killed by the job scheduler for reasons that are not part of the sources of failures summarized above. Additionally, they do not consume compute resources. Table~\ref{tab:features} lists 18 selected features. These features can be categorized into two groups. The first group includes categorical features, such as owner, group, and job name. The other group includes numeric features, such as CPU time usage and integral memory usage. 

When a batch job exits, the scheduler (in our case UGE) generates an \texttt{exit\_status} field in the job accounting data. According to the UGE documentation, a general exit status convention is defined as follows. An exit status of 0 indicates a successful workload. If the command terminates normally, the exit status is the value of the command in the job script, which is in line with normal shell conventions. In the case of the script command exits abnormally, a value of $128$ is added to the value of the command. Thus, the exit status $\geq 128$ can be decomposed into $128 + a\ system\ signal$, where the system signal value can be a fatal error signal such as 6 (\texttt{SIGABRT}), 9 (\texttt{SIGKILL}). 

We summarize the exit status that indicates failure in Table~\ref{tab:exit}. We find that the most common exit status was 1 (58.16\%), indicating that there are miscellaneous errors causing the failures. The next most significant exit status is 7 (17.83\%), which occurs anytime a user feeds too many arguments in the job submission script. We also notice that there are 190 (0.52\%) workloads that are killed by users through system signal 9. Since we do not consider cancelled workloads as failures, we drop these 190 workloads with exit\_status 137. In addition, we do not intent to predict exact errors, so we convert all non-zero exit\_status to 1, representing workloads that face problems during run time. We use exit\_status to distinguish workloads that had completed successfully (exit\_status as 0).

\subsection{Proportion of workload failures}

Figure~\ref{fig:abnormal} presents the proportion and resource consumption characteristics of workload failures. As shown in the figure, workload failure rate is 8.5\% for all submitted jobs (including successful and failed workloads) in quantity. We further analyze the CPU time consumed by failed workloads and find that failed workloads cost 21.1\% of the total CPU time. The proportion of CPU time for failed workloads is larger than the proportion of the number of failed workloads, indicating that the more processors a workload uses and the longer it runs, the higher the probability that this workload will fail. Additionally, we quantify the integral memory usage consumed by failed workloads. As shown in Figure~\ref{fig:abnormal}, the wasted memory resource rate is 20.2\%. All these statistics imply that failed workloads waste significant amount of computational resources and therefore degrade the system efficiency. 

\subsection{Distribution by nodes}

\begin{figure}
     \centering
     \begin{subfigure}[b]{0.48\textwidth}
         \centering
        \includegraphics[width=0.95\linewidth]{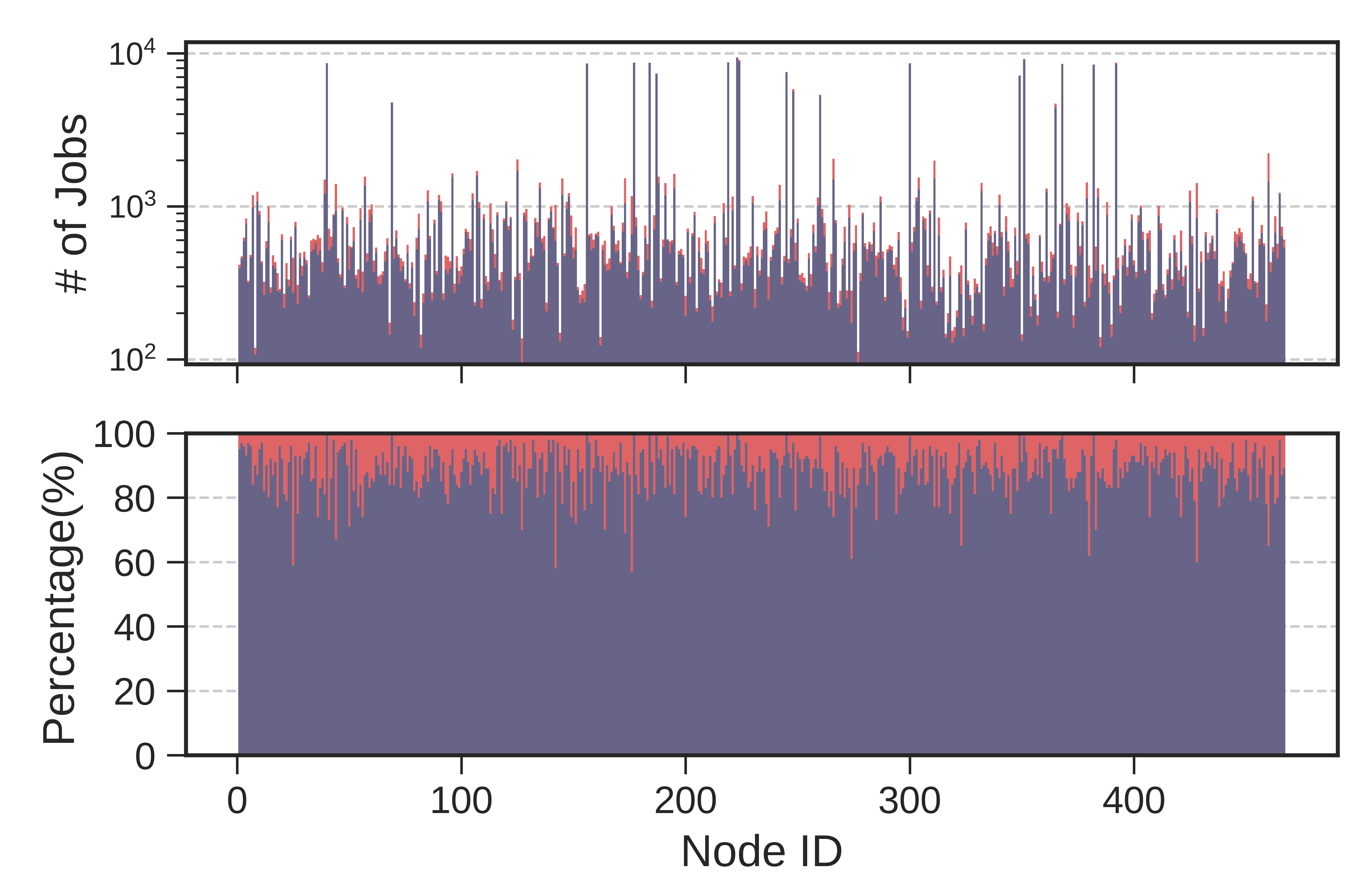}
        \caption{}
    \label{fig:hostname}
     \end{subfigure}
     \begin{subfigure}[b]{0.48\textwidth}
         \centering
        \includegraphics[width=0.95\linewidth]{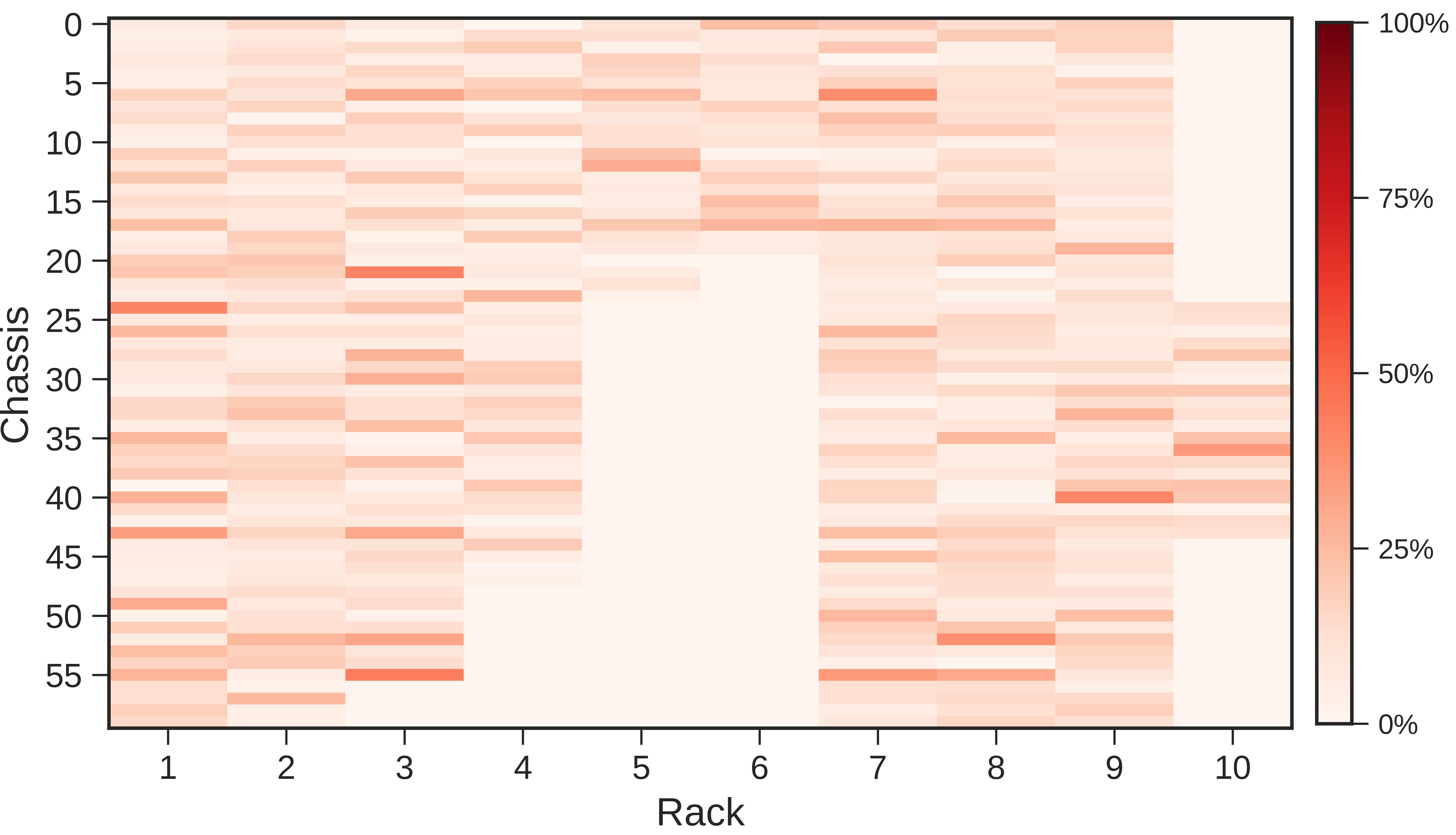}
        \caption{}
    \label{fig:rackslots}
     \end{subfigure}
        \caption{Number and percentage of workload failures per node distributed by node ID (a) and by physical location (b). In sub-figure (a), red indicates failed workloads and dark blue indicates successful workloads. In sub-figure (b), the darkness of the color represents the workload failure rate (in other words, the darker the color means the higher the workload failure rate.}
        \label{fig:jobnodeid}
\end{figure}

We depict the distribution of failures across the nodes of the system by node ID and physical locations in Figure~\ref{fig:hostname} and Figure~\ref{fig:rackslots}, respectively. Figure~\ref{fig:hostname} shows the total number and percentage of workload failures for each node. We first observe that about 20 nodes serve a relatively large number of workloads than the other nodes, while some nodes have more than 60\% of workload failures. It is worth noting that the nodes serving a large number of workloads do not necessarily have a higher percentage of workload failures. A possible explanation is that nodes with high workload failure rate may have node-specific (hardware or operating system) vulnerabilities. 

These 467 nodes in Quanah cluster are hosted in 10 racks and each node can be uniquely addressed by rack and chassis number. Each column shown in Figure~\ref{fig:rackslots} represents one rack of nodes and each row represents one chassis. From Figure~\ref{fig:rackslots}, we observe that nodes in rack 1, 3 and 7 have relatively high workload failure rate. Since the power, temperature and connectivity of all nodes located in a rack are controlled together, problems in these areas can cause failures to occur in physical location vicinity~\cite{ghiasvand2019anomaly}. 

\subsection{Distribution by users}\label{dis_users}

\begin{figure}
     \centering
     \begin{subfigure}[b]{0.48\textwidth}
         \centering
        \includegraphics[width=0.95\linewidth]{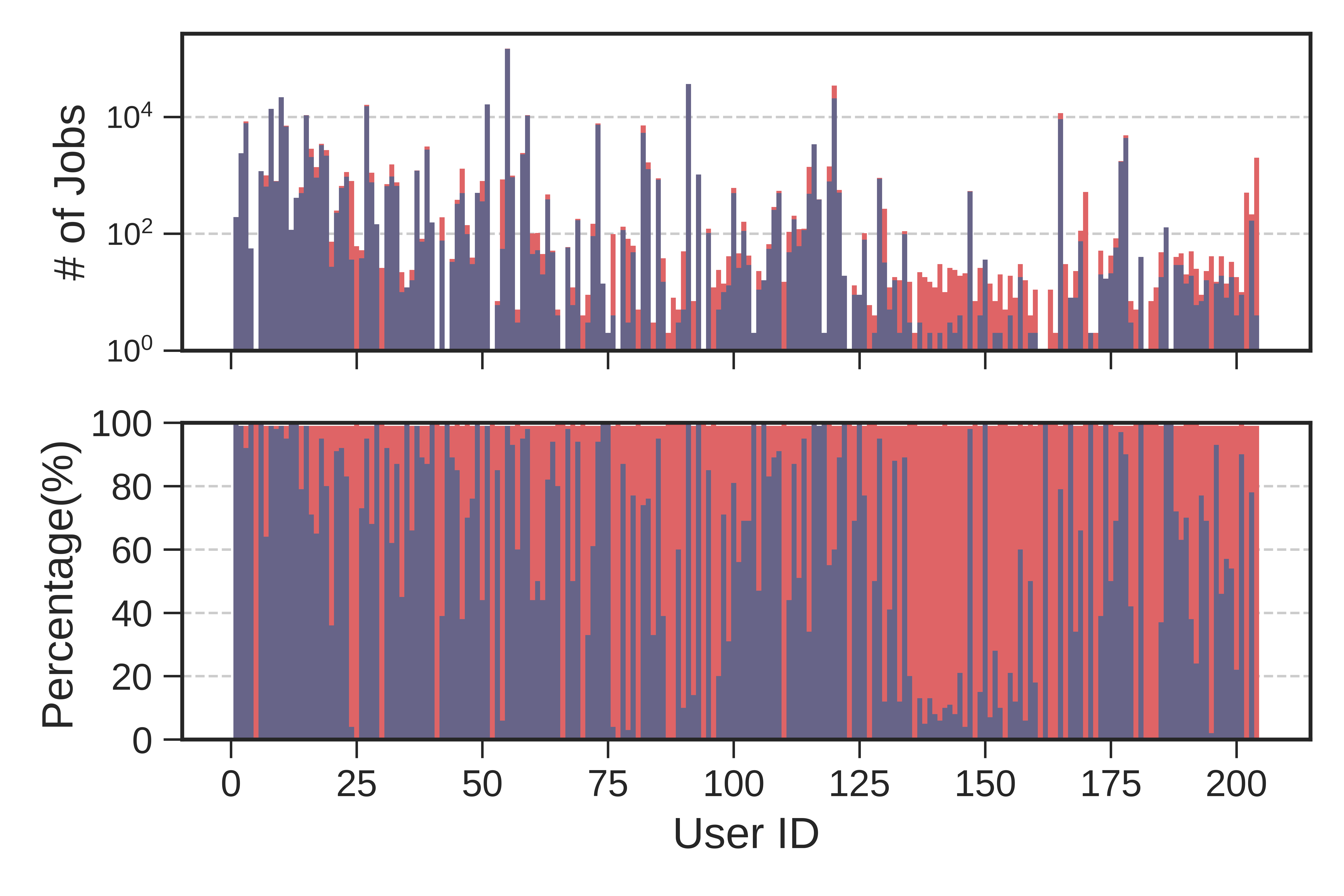}
        \caption{}
    \label{fig:owner}
     \end{subfigure}
     \begin{subfigure}[b]{0.48\textwidth}
         \centering
        \includegraphics[width=0.95\linewidth]{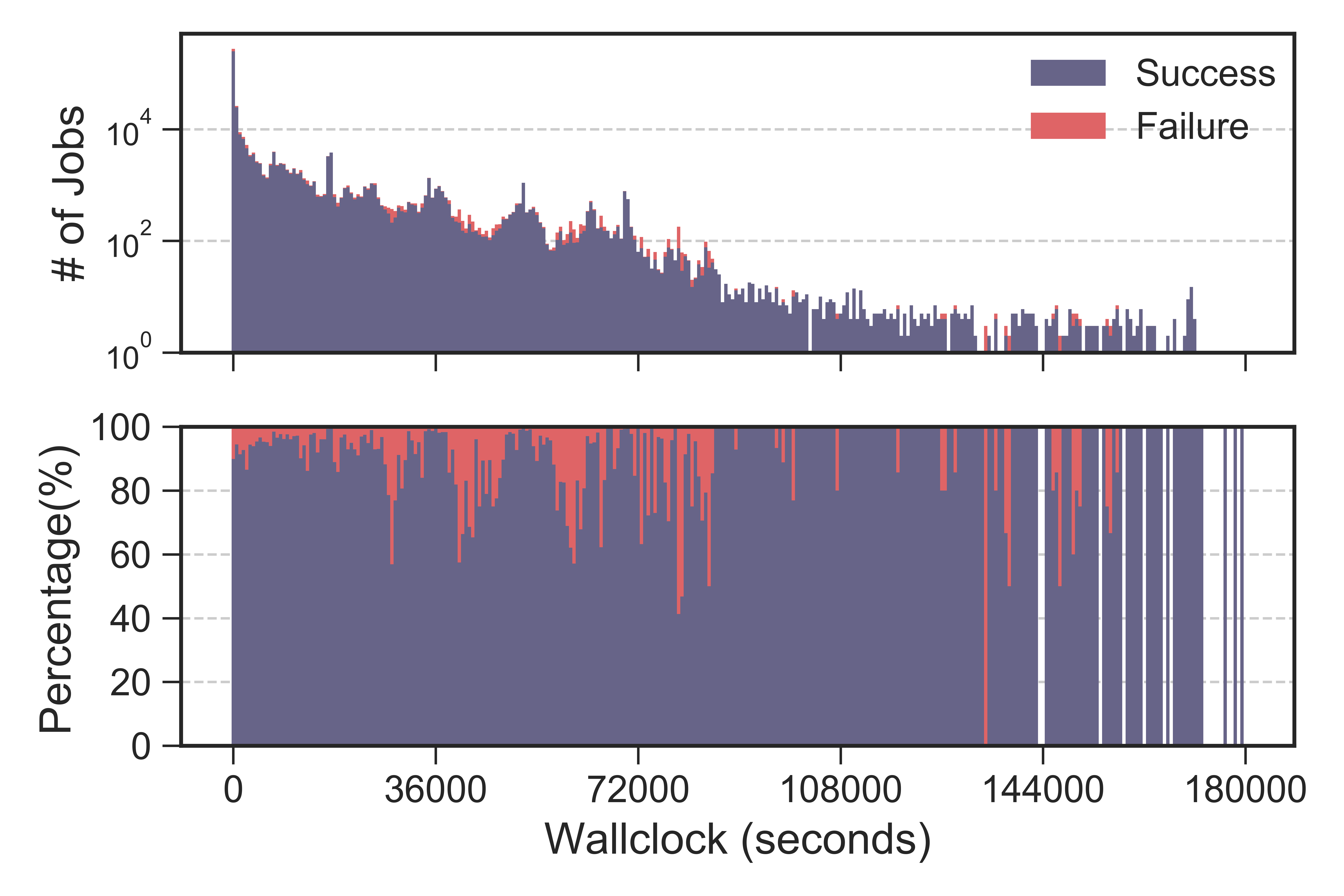}
        \caption{}
        \label{fig:wallclock}
     \end{subfigure}
        \caption{Number and percentage of workload failures distributed by user ID (a) and by wallclock (b).}
        \label{fig:jobownertime}
\end{figure}

To find out the correlation between users and failed workloads, we plot the workload distribution by user ID, as shown in Figure~\ref{fig:owner}. The total number of jobs submitted by users ranges from a few to over 10,000. We observe that the failure rate of workloads per user varies significantly and those users who submit a small number of workloads have a large portion of failed workloads. These statistics suggest that users' experience in properly configuring their applications and/or requesting computational resources varies widely and that these inexperienced users contribute a large fraction of failed workloads.

\subsection{Distribution by time}\label{dis_time}

\begin{figure}
     \centering
     \begin{subfigure}[b]{0.48\textwidth}
         \centering
    \includegraphics[width=0.95\linewidth]{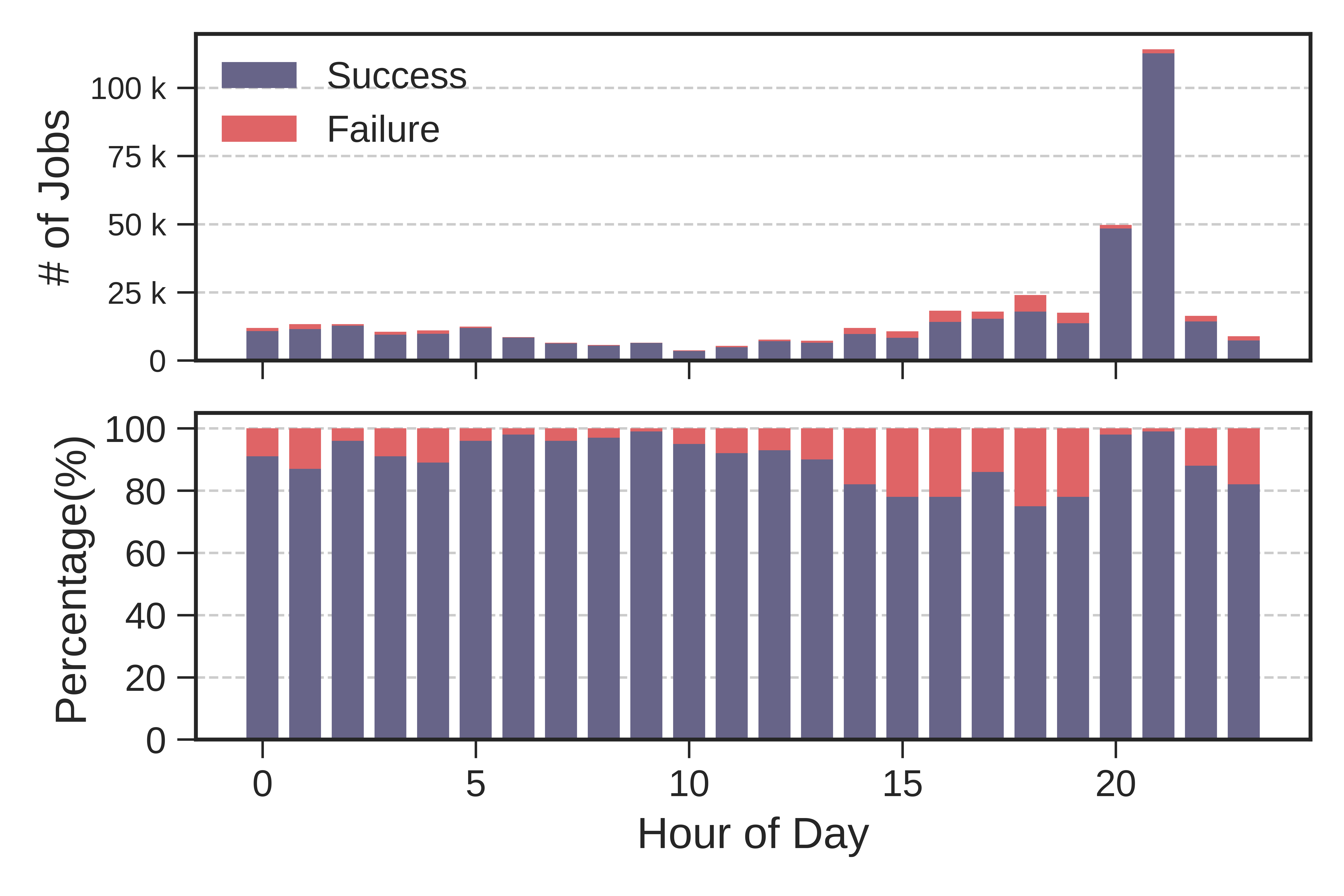}
    \caption{}
    \label{fig:hour}
     \end{subfigure}
     \begin{subfigure}[b]{0.48\textwidth}
         \centering
        \includegraphics[width=0.95\linewidth]{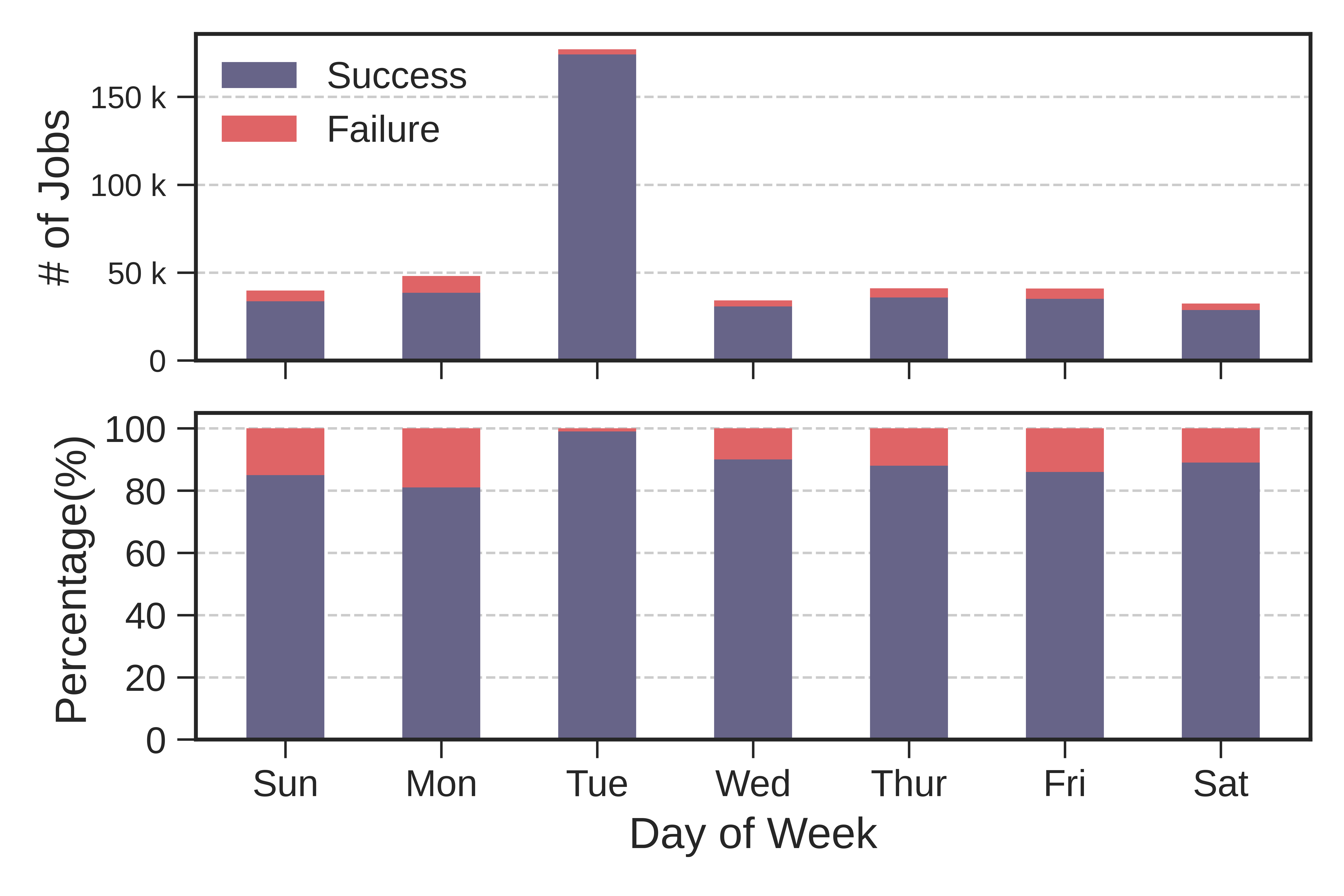}
        \caption{}
        \label{fig:week}
     \end{subfigure}
        \caption{Number and percentage of workload failures distributed by hour of the day (a) and by the day of the week (b).}
        \label{fig:jobtime}
\end{figure}

The wallclock is the actual time taken from the start to the end of a workload. On Quanah cluster, if the user does not specify a runtime in the script, the job scheduler has a default runtime limit of 172,800 seconds (48 hours) for each submitted job. We illustrate the distribution of workloads by wallclock in Figure~\ref{fig:wallclock}. In general, the number of workloads decreases as the wallclock increases, and there is no significant correlation between the failure rate of workloads and the number of workloads. However, we observe a reverse correlation between 24,000s and 84,000s: high workload intensities are associated with low failure rates and vice versa. In addition, the failure rate appears to be high around 144,000s as well.

It is commonly known that the usage pattern of data centers fluctuates with time~\cite{schroeder2009large}. Figure~\ref{fig:jobtime} categorizes the workload failures by the hour of a day and by the day of a week. We observe that there is a slight correlation between failure rate and time. During the time when the highest and the lowest number of jobs occur, the failure rate is below 5\%. However, during the rest of time, the failure rate is between 10\% and 22\%. A possible explanation for this observation is that experienced users submit large array jobs that contribute to the majority of the workloads. Furthermore, their code is robust and their applications are well configured, resulting in low failure rates. When the workload intensity is low, the system is less vulnerable. The failure rate for each day of a week shows similar results: the highest workload volumes resulted in the lowest failure rate. However, during the rest of days of a week, the workload failure rate does not change much. 
\section{Methodology}\label{models}

\begin{figure}
\centering
\includegraphics[width=0.98\linewidth]{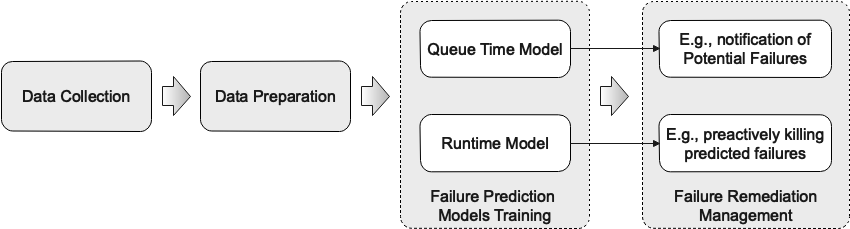}
\caption{Workflow of Predicting Workload Failures in Data Centers}
\label{fig:workflow}
\end{figure}

In this section, we describe the workflow of predicting workload failures in data centers. As shown in Figure~\ref{fig:workflow}, the workflow consists of four phases: (1) data collection: collecting metrics from data centers; (2) data preparation: preprocessing the data into a structured format and extracting features for machine learning models; (3) model training: training \emph{Queue-time} and \emph{Runtime} models using machine learning algorithms; (4) remediation management: applying remediation management techniques to leverage the prediction results to optimize the management of data centers. Data collection has already been discussed in Section~\ref{monitoring}. Therefore, we focus on data preparation and model training in this research. We leave the failure remediation management and data center optimizations as near-future research work.

\subsection{Data Preparation}

\subsubsection{Preprocessing}
In our current design and implementation, the job accounting data is stored in a MySQL database; we perform a select operation with start and end times to select the data collected from August 1th, 2020 to October 1th, 2020. The data is then saved into a dataframe. In the data preprocessing phase, we convert raw features into a format more suitable for machine learning training. Specifically, we create dummy variables for the categorical variables by using one-hot encoding~\cite{dummy}, and scale the numerical features to avoid features with high variability from having more influence in the prediction.

Another important design consideration in data preparation is to deduct irrelevant attributes and derive appropriate features from the original attributes. Irrelevant attributes add extra dimensions to the data set and can distract machine learning algorithms from achieving accurate prediction rules. On the other hand, deriving proper combinational features can boost the prediction accuracy. In our case, we drop the feature \emph{job\_id} because it is assigned by the job scheduler and does not reveal the characteristics of the workload. We derive hours of the day and days of the week from time-related features to augment the data. We also derive several numeric features from the resource usage data, such as CPU intensity and average memory usage. CPU intensity is defined as ${(cpu/slots)/wallclock }$, i.e., the ratio of the CPU time of a workload's single processor to its overall wallclock (i.e. runtime). The average memory usage is defined as ${mem/wallclock}$, i.e., the ratio of the integral memory usage of a workload to its run time.

In addition, the collected data does not contain information about applications and libraries. To overcome this limitation, we apply Natural Language Processing (NLP) techniques to job names and identify similar job names submitted by the same user. We then assign a uniform name to these workloads as their job names. This process aims to categorize workloads that use the same libraries. An underline hypothesis of this process is that workloads with similar job names submitted by the same user tend to be the same applications and use the same libraries, differing only in parameters or parts of the code.

As discussed in Section~\ref{workload}, we do not intend to predict the exact workload error, therefore the prediction is a binary classification problem (i.e., success or failure). We convert exit\_status to 1 if the workload fails and 0 otherwise, and use this as the class label. 

\subsubsection{Features Selection}\label{features}

Predicting workload failures provides input for failure remediation management, where possible techniques include: 1) notifying users of potential workload failures after job submission but before execution; 2) making better scheduling decisions based on the prediction, thereby encouraging users to improve code quality and request appropriate computational resources; and 3) killing workloads that will fail before wasting too many computational resources. To this end, we plan to train two models, one for predicting pre-run failures (i.e., queue-time model) and the other for predicting runtime failures (i.e., runtime model). 

Queue-time model and runtime model are trained with features available in different job states. The queue-time model is trained with categorical features such as owner, group, job name, department, etc. The runtime model uses not only categorical features but also resource usage features, such as CPU time, integral memory usage, data transferred in IO, etc. Note that in our case, resource request data, such as estimated job running time and expected maximum memory needed during runtime, are not recorded in the job accounting data, slightly limiting the available features that can be used to train the queue-time model. For other data sets that include resource request information, the queue-time model can be enhanced and its prediction results should be more accurate.  

\subsection{Model Training}
We use five classification algorithms to implement machine learning models and train them with our 324,358 instances to predict workload failures. These algorithms and the corresponding hyper-parameters are described below.

\textbf{Gaussian Naive Bayes}: Naive Bayes is a probabilistic machine learning algorithm based on the Bayes Theorem, which is a simple mathematical formula for calculating conditional probabilities. In our implementation, we use Gaussian Naive Bayes (GNB) (i.e., Naive Bayes extended to real-valued attributes). It is easy to implement because we only need to estimate the mean and standard deviations of the training data. Guassian NB does not accept parameters, except for the \texttt{priors} parameter, which we use the default value of ``None'' in our model.

\textbf{Logistic Regression}: Logistic Regression (LR) is a classification algorithm for finding the relationship between features and outcome probabilities and is the most widely used machine learning algorithm in classification problems. It is relatively fast compared to other supervised classification techniques. Since we do not predict the exact value of the exit status, we use Binomial Logistic Regression. Logistic Regression does not actually have any critical hyper-parameters to tune. We set the inverse regularization parameter (i.e. \texttt{C}) to 0.1 and choose ``l2'' as the \texttt{penalty} parameter and ``liblinear'' as the \texttt{solver} parameter.

\textbf{Linear Discriminant Analysis}: Linear Discriminant Analysis (LDA), as the name implies, is most commonly used as a dimensionality reduction technique, but it can also be used as a classification tool by finding linear combinations of features that separate two or more classes. LDA works by calculating summary statistics, such as mean and standard deviation, of input features by class label. Predictions are performed by estimating the probability that a new instance belongs to each class label based on the values of each feature. We set the \texttt{solver} to ``lsqr'', which performs best in our data set compared to other built-in solvers.

\textbf{Decision Tree}: Decision Tree (DT) is a predictive model that predicts value by learning decision rules inferred from data features. One of the advantages of this algorithm is that the non-linear relationship between features does not affect the performance of the tree. It can handle both categorical and numeric data. The \texttt{criterion} parameter in the DT is set to ``gini'' and the \texttt{splitting} parameter is set to ``best''. All other parameters are kept as default. 

\textbf{Random Forest}: Random Forest (RF) is an ensemble method that consists of a large number of individual decision trees. It uses bagging and feature randomness in the construction of each tree to create a forest of uncorrelated trees. Each individual tree in the random forest produces a class prediction and the class with the most votes will be the predicted value of the model. As with the Decision Tree, we set the \texttt{criterion} parameter to ``gini'' instead of ``entrophy''. The number of random features (i.e. \texttt{max\_features}) considered in each split is set to ``sqrt'', which is usually good for classification problems. The rest of the parameters are left unchanged.

Since our data set is very large, we use the holdout method instead of the cross-validation method to save computational cost. The data set is partitioned into 65\% training data, 15\% validation data and 20\% testing data. The training set learns the relationship between the features and the target variables (i.e. 0 for success and 1 for failure). The validation set is used to check how accurately the model defines the relationship between features and known outcomes. The testing data provides a final estimate of the model performance after the model has been trained and validated.

\section{Experimental Results}\label{results}

In this section, we describe the evaluation metrics we used for our experiments and present the experimental results including the performance of the machine learning algorithms described above and the potential resource savings that benefit from the prediction, followed by an evaluation of the impact of the training sizes. The models in this study are implemented in the \emph{scikit-learn}~\cite{pedregosa2011scikit} Python library.

\subsection{Evaluation Metrics}

\subsubsection{Prediction Metrics}

In order to measure the performance of ML algorithms, it is important to specify evaluation metrics. We use \emph{recall} (i.e., true positive rate), \emph{precision} and \emph{F1 Score} as our measurements. Recall represents the ratio between the number of correctly predicted failed workloads to  the total number of actual failures. Precision is calculated by dividing the total number of predicted failures with the number of correctly predicted failed workloads. F1 score is the weighted average of recall and precision. A higher score for these three metrics means that the model's classification results are more accurate. These measurements are shown below:

\begin{equation}\label{eq:recall}
    recall = \frac{\#\ of\ Correctly\ Predicted\ Failures}{Total\ \#\ of\ Actual\ Failures}
\end{equation}

\begin{equation}\label{eq:prediction}
    precision = \frac{\#\ of\ Correctly\ Predicted\ Failures}{Total\ \#\ of\ Predicted\ Failures}
\end{equation}

\begin{equation}\label{eq:F1}
    F1\ Score = \frac{2 * (recall * precision)}{recall + precision}
\end{equation}

\subsubsection{Resource Savings Metrics}

The basic proactive failure remediation management is to simply kill workloads that are predicted to fail. This strategy is sensitive to false positive, where workloads are incorrectly predicted to fail. Killing workloads inappropriately will result in wasted resources, as the killed workloads will be restarted and run at a later time. Therefore, We define the resource saving (${R_{saving}}$) as: 

\begin{equation}
    {R_{saving}} = \frac{R_{s} - R_{w}}{{R_{total}}},
\end{equation}

\noindent where ${R_{total}}$ is the total resources consumed by failed and successful workloads, ${R_{s}}$ is the resource saved by proactively killing failed workloads, and ${R_{w}}$ is the resource wasted by killing successful workloads.

\subsection{Failure Prediction}


Table~\ref{table:prerunning} presents the performance of the queue-time model. Specifically, we observe that Gaussian Naive Bayes (GNB) achieves the highest recall score of 99.44\%; Random Forest (RF) performs the best with a precision score of 90.61\% and an F1 score of 87.71\%. The performance of the runtime model are shown in Table~\ref{table:running}. Again, RF achieves the best performance with a precision of 97.75\% and an F1 score of 95.91\%. Although GNB achieves the highest recall score, its precision score is the lowest, indicating a low number of successful failure predictions in its total failure predictions and it predicts most successful workloads as failures. When evaluating the overall performance, we choose RF as the classification algorithm for both models.

\begin{table}[h]
    \centering
    \caption{Performance of Queue-Time Model}
    \label{table:prerunning}
    \begin{tabular}{c c c c c}
        \specialrule{.1em}{.05em}{.05em}
        \rowcolor{Gray}
        \textbf{Model} & \textbf{recall}& \textbf{precision} & \textbf{F1 Score} & \textbf{Training Time(s)}\\
        \specialrule{.1em}{.05em}{.05em}
        GNB & \textbf{99.44} & 15.65 & 27.04 & 3.5 \\
        LR & 57.22 & 86.16 & 68.77 & 5.46 \\
        LDA & 62.14 & 77.92 & 69.14 & 45.12 \\
        DT & 84.02 & 90.33 & 87.06 & 123.96 \\
        RF & 85.00 & \textbf{90.61} & \textbf{87.71} & 149.81 \\
        
        \hline
    \end{tabular}
\end{table}

\begin{table}[h]
    \centering
    \caption{Performance of Runtime Model}
    \label{table:running}
    \begin{tabular}{c c c c c}
        \specialrule{.1em}{.05em}{.05em}
        \rowcolor{Gray}
        \textbf{Model} & \textbf{recall}& \textbf{precision} & \textbf{F1 Score} & \textbf{Training Time(s)}\\
        \specialrule{.1em}{.05em}{.05em}
        GNB & \textbf{99.44} & 15.65 & 27.05 & 5.13 \\
        LR & 58.13 & 86.48 & 69.52 & 6.32 \\
        LDA & 58.92 & 81.64 & 68.45 & 46.7 \\
        DT & 93.92 & 94.57 & 94.24 & 173.08 \\
        RF & 94.14 & \textbf{97.75} & \textbf{95.91} & 145.71 \\
        \hline
    \end{tabular}
\end{table}

\subsection{Resource Savings}

\begin{figure}
    \centering
    \begin{subfigure}[b]{0.45\textwidth}
        \centering
        \includegraphics[width=0.95\linewidth]{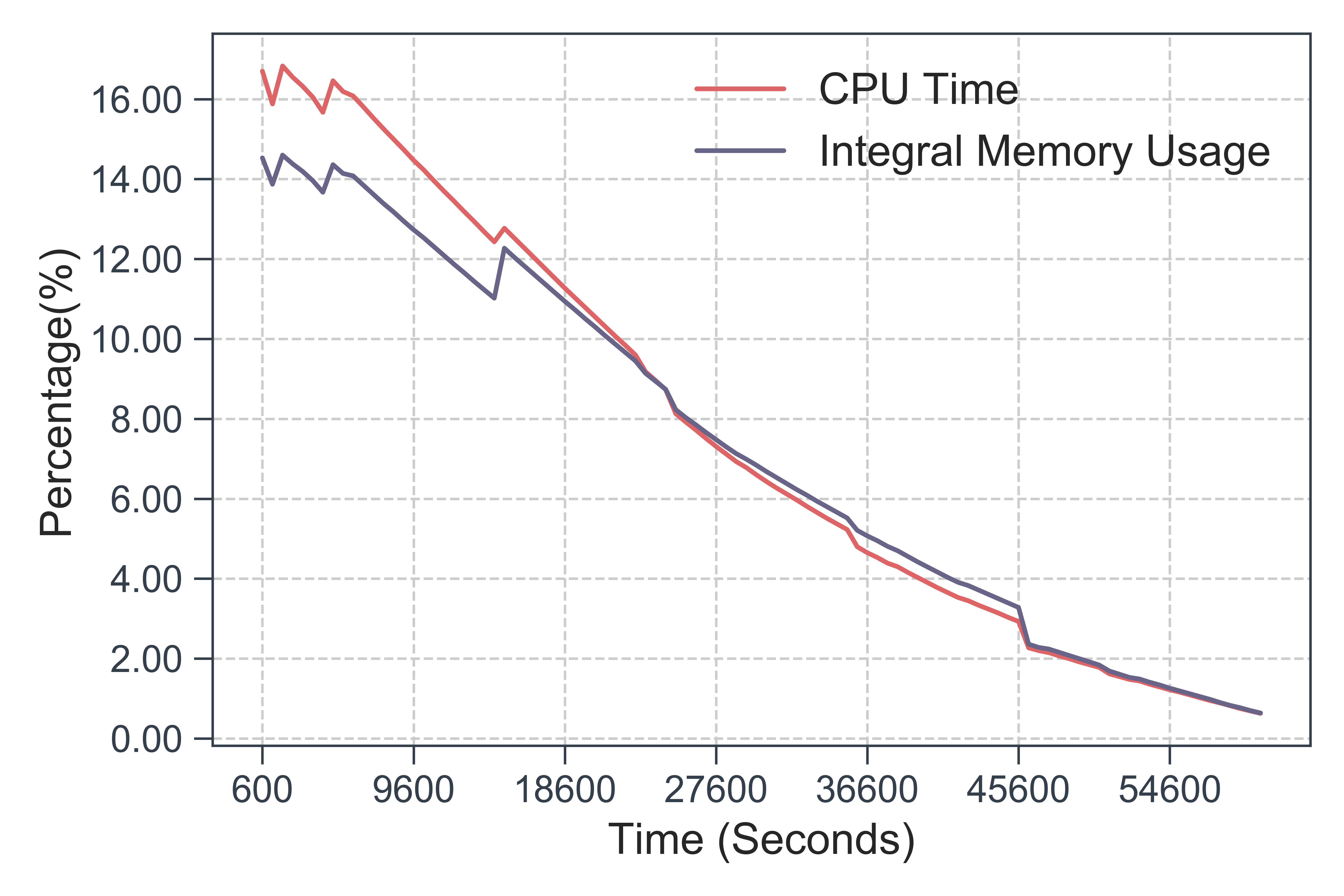}
        \caption{}
        \label{fig:savings}
    \end{subfigure}
    \begin{subfigure}[b]{0.45\textwidth}
        \centering
        \includegraphics[width=0.95\linewidth]{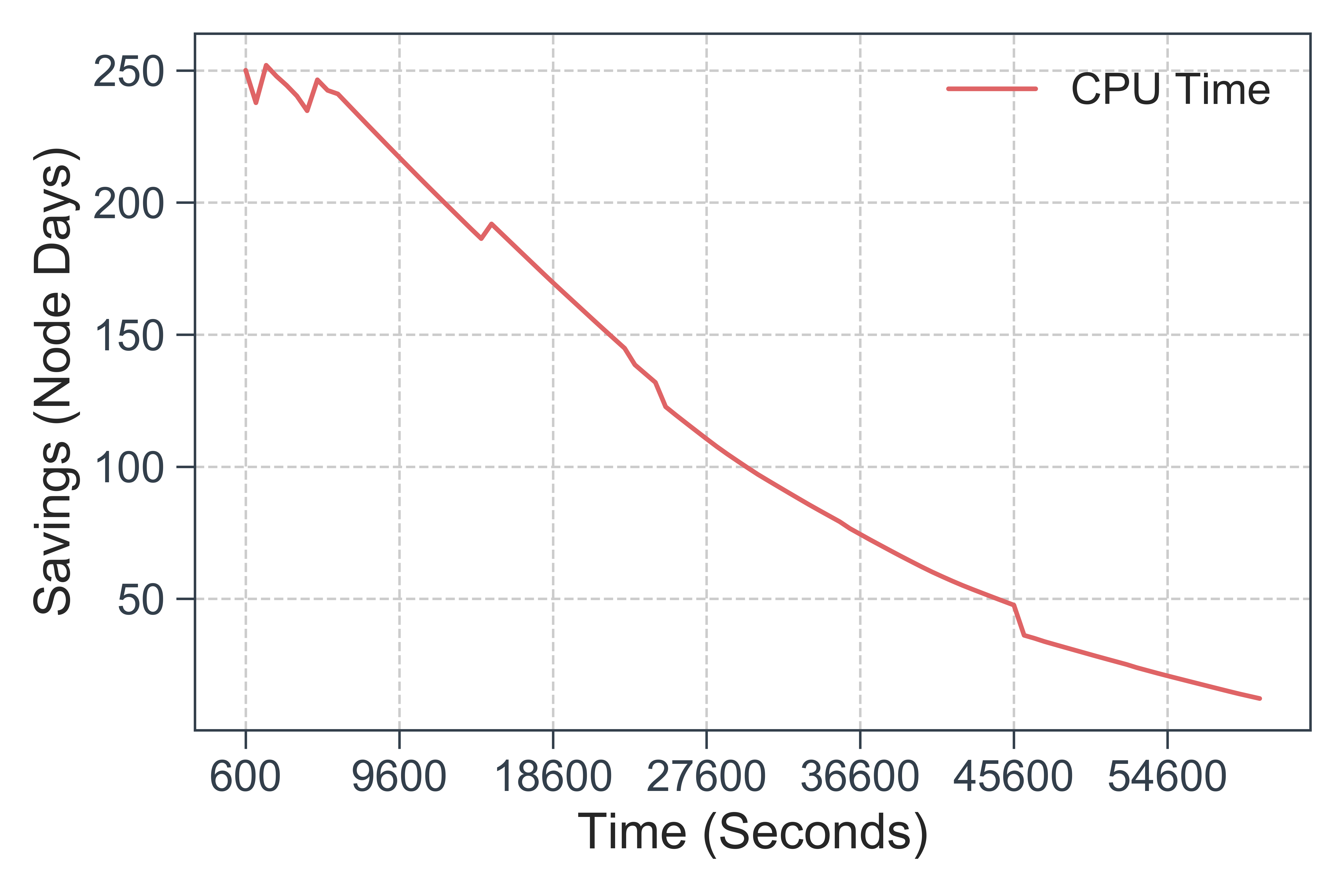}
        \caption{}
        \label{fig:cputime}
    \end{subfigure}
    \caption{Resource Savings in percentage (a) and CPU Time Savings in Node Days (b) at different times.}
    \label{fig:results1}
\end{figure}

The numeric features such as CPU time and integral memory usage are only known after the workload has completed execution. This fact may raise the question of how we can estimate resource savings at different run times of a workload and use these features to predict workload failures early. To apply the run time model on a running workload, we make the following assumption: resource usage is linearly proportional to run time, so its resource usage at different times can be calculated as: 

\begin{equation}
  CRU = FRU * \frac{Time}{Wallclock},
\end{equation}

\noindent where \emph{CRU} stands for Current Resource Usage and \emph{FRU} stands for Final Resource Usage. Based on this formula, we generate a series of test data sets from the original test data (20\% of 2-month workload traces in Quanah cluster, i.e., 12-day workload traces) containing synthetic resource usage at different times, and then, apply the runtime model on these data sets. Figure~\ref{fig:savings} shows the resource savings. From this figure, we observe the same pattern of savings in CPU time and integral memory usage; they both achieve the highest savings at the beginning of the time, 16.7\% and 14.53\%, respectively. Overall, the resource savings decrease over time, except for a few ups and downs at around 4200s and 14400s. To understand the resource savings, we convert the CPU time savings in $seconds$ to CPU time savings in $node \cdotp days$, where the node has 36 CPUs. As shown in Figure~\ref{fig:cputime}, the maximum CPU time savings is about 250 $node \cdotp days$. In other words, applying the runtime model on 12-day workloads of a 467-node cluster will help save the CPU time (and associated power consumption) of a node running for 250 days.

\begin{figure}
    \centering
    \begin{subfigure}[b]{0.45\textwidth}
        \centering
        \includegraphics[width=0.95\linewidth]{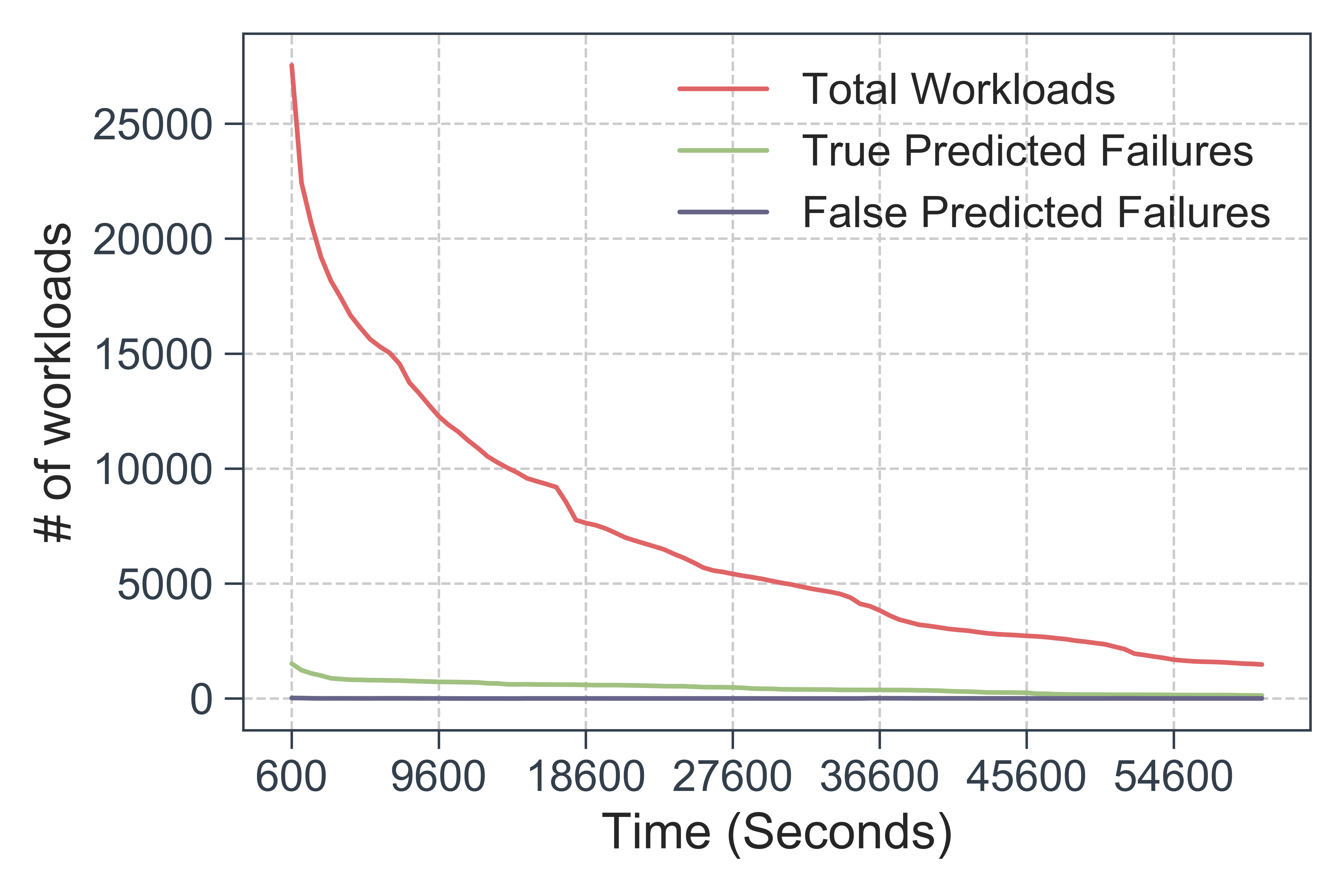}
        \caption{}
        \label{fig:counts}
    \end{subfigure}
    \begin{subfigure}[b]{0.45\textwidth}
        \centering
        \includegraphics[width=0.95\linewidth]{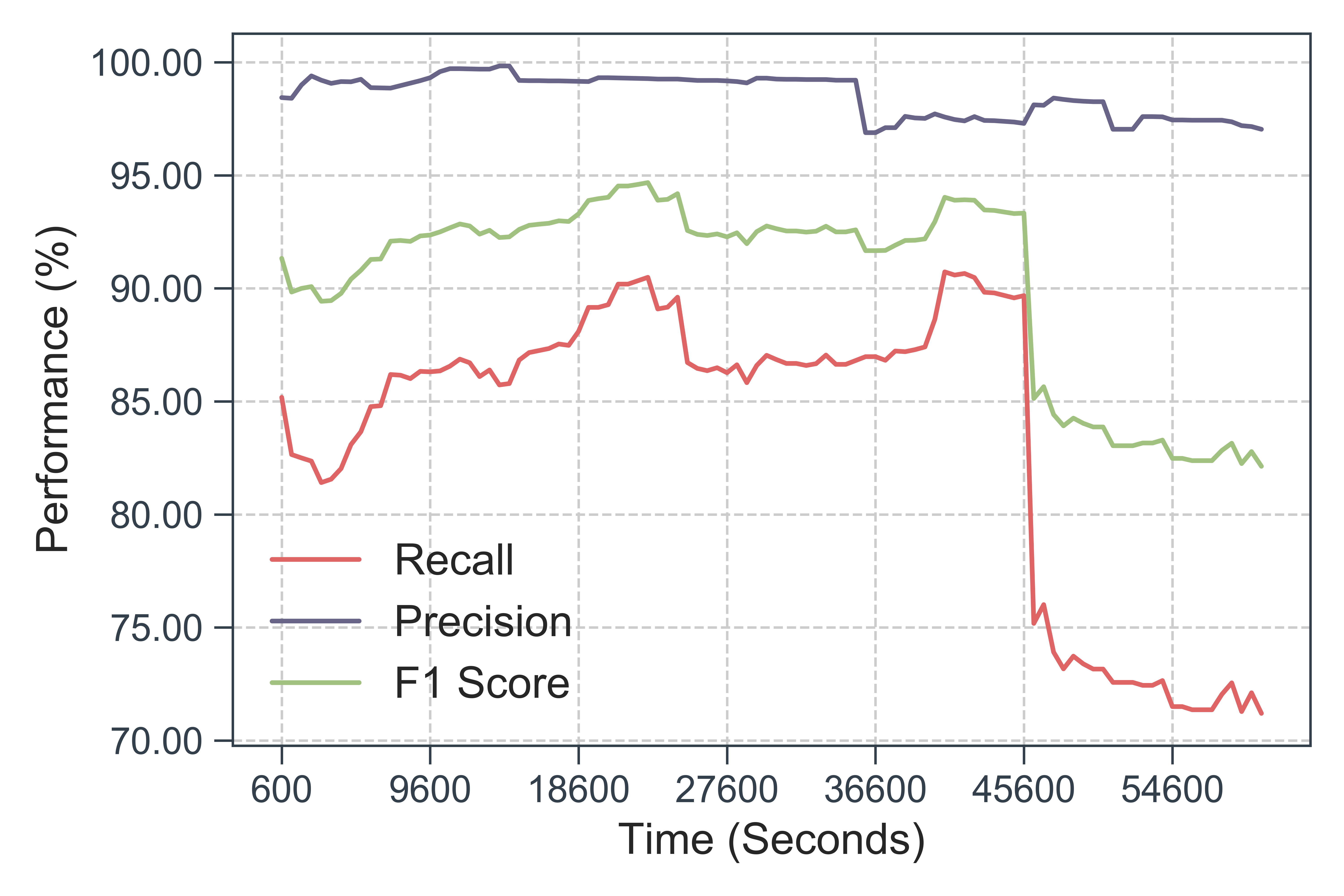}
        \caption{}
        \label{fig:performance}
    \end{subfigure}
    \caption{Number of workloads (a) and Prediction Performance (b) at different times.}
    \label{fig:results2}
\end{figure}

To better understand the resource savings, we plot the number of workloads and prediction performance at different times, as shown in Figure~\ref{fig:counts} and Figure~\ref{fig:performance}. Figure~\ref{fig:counts} shows that the total number of workloads participating in the failure remediation management decreases exponentially and some workloads complete before the runtime model is applied. Therefore, the resource savings that can be achieved decrease with time. Figure~\ref{fig:performance} presents the recall, precision, and F1 scores. Throughout time, the precision scores are at high values. The recall and F1 scores are decent (both scores are above 72\%) although some fluctuations exist. 

\subsection{Effect of Training Size}
\begin{figure}
    \centering
    \begin{subfigure}[b]{0.48\textwidth}
        \centering
        \includegraphics[width=0.95\linewidth]{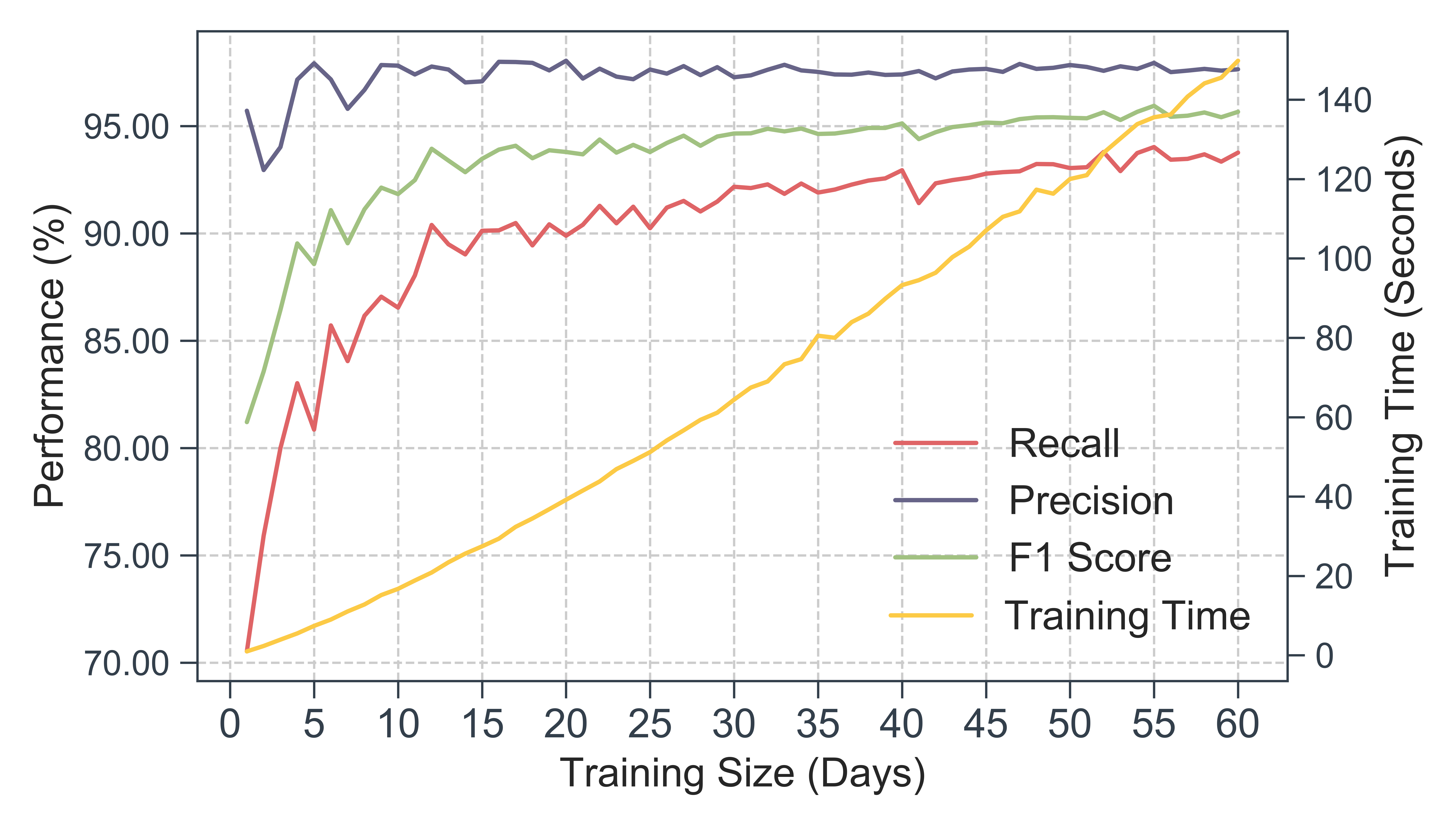}
        \caption{}
        \label{fig:training_size_queue}
    \end{subfigure}
    \begin{subfigure}[b]{0.48\textwidth}
        \centering
        \includegraphics[width=0.95\linewidth]{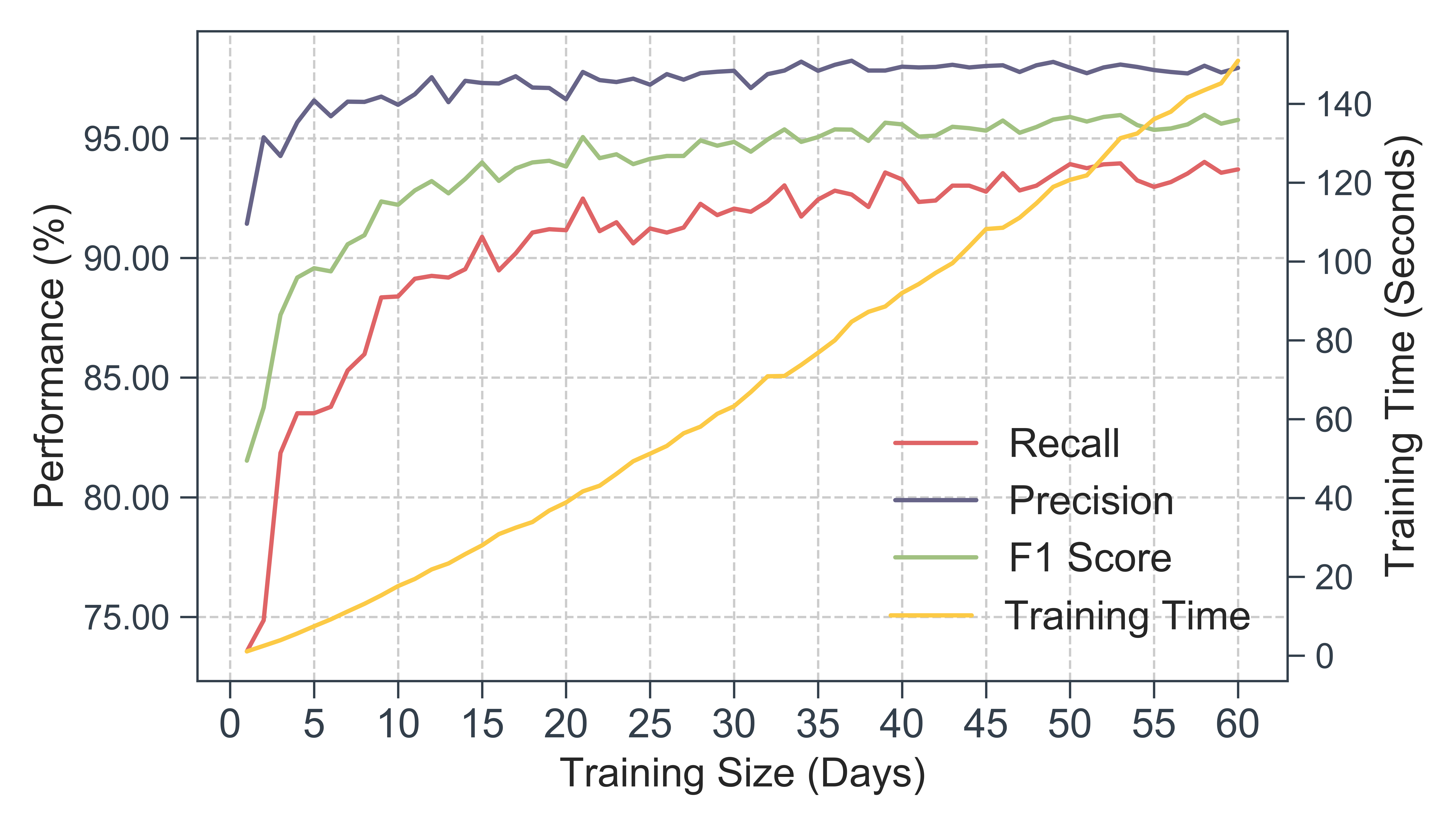}
        \caption{}
        \label{fig:training_size_run}
    \end{subfigure}
    \caption{Training Size vs. Prediction Performance of Queue-time Model (a) and Runtime Model (b).}
    \label{fig:results3}
\end{figure}

Even though we achieve a promising prediction accuracy using Random Forest, the training time is long because of using a large amount of workload traces. As shown in Table~\ref{table:prerunning} and Table~\ref{table:running}, the training time in both models are about 150s. In order to find the optimum training size that achieves a balance between prediction accuracy and training time, we build the prediction models using different training sizes in the range of 1 day to 60 days of data. %
As shown in Figure~\ref{fig:training_size_queue} and Figure~\ref{fig:training_size_run}, the precision, recall and F1 scores of the queue-time model and the runtime model do not improve significantly after the training size exceeding 30 days of data. With the training size of 30 days of data, the training time shortens to 67 seconds, which is acceptable to many data centers to conduct workload failure predictions periodically. 

\section{Related Work}\label{related}

Characterizing and quantifying failures in data centers are invaluable for system administrators to understand the behavior of the systems and thus develop strategies to improve the RAS of the systems. Many prior works have investigated failures on large-scale systems~\cite{fadishei2009job, schroeder2009large, zheng2011co, el2013reading, di2014lessons, ghiasvand2016lessons, ghiasvand2019anomaly}. For example, Fadishei et al.~\cite{fadishei2009job} analyzed workload traces in grid environment and discovered correlations between failure characteristics and performance metrics. Schroeder et al.~\cite{schroeder2009large} examined statistics on failure data collected at two large HPC sites and discovered temporal and spatial correlations of failures. Zheng et al.~\cite{zheng2011co} presented a co-analysis of RAS and job logs that helps in understanding failure patterns and system/user behavior. There are also studies that looked specifically into the reliability of particular component such as DARMs, disks and GPUs~\cite{hwang2012cosmic, sridharan2013feng, nie2017characterizing}.

Considering the failure characteristics and the correlations between failures and job types, performance metrics and components, several studies investigated machine learning models to predict failures on large-scale systems~\cite{fu2012hybrid, chen2014failure, islam2017predicting, nie2018machine}. Fu et al.~\cite{fu2012hybrid} proposed a hybrid failure detection framework using one-class and two-class support vector machines (SVM). Chen et al.~\cite{chen2014failure} proposed a prediction method based on Recurrent Neural Network (RNN) that predicts application failure in cloud using the Google cluster workload traces. Tariqul et al.~\cite{islam2017predicting} developed a similar approach like Chen's by using Long Short-Term Memory Network (LSTM). Many of the proposed approaches are limited to certain performance metrics, such as studies based on Google cluster workload traces~\cite{chen2014failure, islam2017predicting}, or are limited to certain components of the system, such as studies focused on GPUs~\cite{nie2018machine}.

The drawback of these mentioned approaches is that they ignore the human factors that lead to failures. As shown in Section~\ref{dis_users} and Section~\ref{dis_time}, there are correlations between failures and user behavior. A well-trained and experienced user can potentially produce less failure jobs. The proposed approach in this work considers not only performance metrics, but also user behavior in the prediction models. In addition, the proposed approach does not rely on complex system logs collection and analysis; it utilizes job accounting data that is available in all resource managers. Therefore, the prediction models and failure remediation mechanisms (e.g. killing predicted failures) are easier to integrate into resource managers.

\section{Conclusions and Future Work}\label{conclusion}
In this study, we have analyzed two months of job accounting data collected from a production data center and found that failed workloads accounted for 8.5\% of total workloads, consumed 21.1\% of the total CPU time and 20.2\% of the integral memory usage. In addition, we have quantified the workload failure rates across nodes, users, and different time scales, and we have analyzed the correlation between them. Based on the comprehensive understanding of workload traces, we develop two prediction models (queue-time model and runtime model) with five machine learning algorithms and have found that Random Forest performed the best with 90.61\% and 97.75\% precision scores, respectively. We further explored the training size and its impact on prediction performance and training time, and we concluded that 30 days of job data is the optimal training size, with 67 seconds of training time for our data sets. Our experimental results show that the workload failure prediction model can help save CPU time and integrated memory usage by up to 16.7\% and 14.53\%, respectively. 

Nevertheless, our study can be further improved in several aspects. First, due to the lack of resource usage data for workloads at different runtimes, we had to create synthetic data to quantify the resource savings gained from the runtime model. This approach may not be representative of all situations and the accuracy of the predictions may not be as high as expected. Second, because resource request information is an important factor in predicting workload failure, the lack of this feature prevented our model from achieving more accurate predictions. Third, the prediction models only predict the probability of workload failure. Even though we have achieved promising performance, we cannot infer the causality of workload failure based on the available data since correlation does not imply causality. To further support causality identification, we plan to develop a provenance based approach for failure predictions in the future.

In large-scale data centers, where workload failures become the norm, proactive failure management is critical to improve system reliability, availability, and scalability. In future work, we plan to improve the prediction by adding more features in the training data, such as hardware monitoring metrics and system logs, and explore other machine learning algorithms, such as LSTM. In addition, understanding the causality of workload failures is important for both system administrators and users. We hope to conduct causal inference studies when the detailed provenance is available. Moreover, failure-aware resource scheduling is also a promising research direction and deserves further studies.

\balance
\bibliographystyle{IEEEtran}
\bibliography{ref}

\begin{thebibliography}{10}
\providecommand{\url}[1]{#1}
\csname url@samestyle\endcsname
\providecommand{\newblock}{\relax}
\providecommand{\bibinfo}[2]{#2}
\providecommand{\BIBentrySTDinterwordspacing}{\spaceskip=0pt\relax}
\providecommand{\BIBentryALTinterwordstretchfactor}{4}
\providecommand{\BIBentryALTinterwordspacing}{\spaceskip=\fontdimen2\font plus
\BIBentryALTinterwordstretchfactor\fontdimen3\font minus
  \fontdimen4\font\relax}
\providecommand{\BIBforeignlanguage}[2]{{%
\expandafter\ifx\csname l@#1\endcsname\relax
\typeout{** WARNING: IEEEtran.bst: No hyphenation pattern has been}%
\typeout{** loaded for the language `#1'. Using the pattern for}%
\typeout{** the default language instead.}%
\else
\language=\csname l@#1\endcsname
\fi
#2}}
\providecommand{\BIBdecl}{\relax}
\BIBdecl

\bibitem{cappello2014toward}
F.~Cappello, G.~Al, W.~Gropp, S.~Kale, B.~Kramer, and M.~Snir, ``Toward
  exascale resilience: 2014 update,'' \emph{Supercomputing Frontiers and
  Innovations: an International Journal}, vol.~1, no.~1, pp. 5--28, 2014.

\bibitem{candea2004recovery}
G.~Candea, A.~B. Brown, A.~Fox, and D.~Patterson, ``Recovery-oriented
  computing: Building multitier dependability,'' \emph{Computer}, vol.~37,
  no.~11, pp. 60--67, 2004.

\bibitem{candea2004microreboot}
G.~Candea, S.~Kawamoto, Y.~Fujiki, G.~Friedman, and A.~Fox, ``Microreboot--a
  technique for cheap recovery,'' \emph{arXiv preprint cs/0406005}, 2004.

\bibitem{hargrove2006berkeley}
P.~H. Hargrove and J.~C. Duell, ``Berkeley lab checkpoint/restart (blcr) for
  linux clusters,'' in \emph{Journal of Physics: Conference Series}, vol.~46,
  no.~1.\hskip 1em plus 0.5em minus 0.4em\relax IOP Publishing, 2006, p. 067.

\bibitem{garg2011environment}
S.~K. Garg, C.~S. Yeo, A.~Anandasivam, and R.~Buyya, ``Environment-conscious
  scheduling of hpc applications on distributed cloud-oriented data centers,''
  \emph{Journal of Parallel and Distributed Computing}, vol.~71, no.~6, pp.
  732--749, 2011.

\bibitem{aupy2016co}
G.~Aupy, M.~Shantharam, A.~Benoit, Y.~Robert, and P.~Raghavan, ``Co-scheduling
  algorithms for high-throughput workload execution,'' \emph{Journal of
  Scheduling}, vol.~19, no.~6, pp. 627--640, 2016.

\bibitem{rodriguez2019job}
M.~Rodr{\'\i}guez-Pascual, J.~Cao, J.~A. Mor{\'\i}{\~n}igo, G.~Cooperman, and
  R.~Mayo-Garc{\'\i}a, ``Job migration in hpc clusters by means of
  checkpoint/restart,'' \emph{The Journal of Supercomputing}, vol.~75, no.~10,
  pp. 6517--6541, 2019.

\bibitem{garg2018shiraz}
R.~Garg, T.~Patel, G.~Cooperman, and D.~Tiwari, ``Shiraz: Exploiting system
  reliability and application resilience characteristics to improve large scale
  system throughput,'' in \emph{2018 48th Annual IEEE/IFIP International
  Conference on Dependable Systems and Networks (DSN)}.\hskip 1em plus 0.5em
  minus 0.4em\relax IEEE, 2018, pp. 83--94.

\bibitem{elnozahy2004checkpointing}
E.~N. Elnozahy and J.~S. Plank, ``Checkpointing for peta-scale systems: A look
  into the future of practical rollback-recovery,'' \emph{IEEE Transactions on
  Dependable and Secure Computing}, vol.~1, no.~2, pp. 97--108, 2004.

\bibitem{cappello2009fault}
F.~Cappello, ``Fault tolerance in petascale/exascale systems: Current
  knowledge, challenges and research opportunities,'' \emph{The International
  Journal of High Performance Computing Applications}, vol.~23, no.~3, pp.
  212--226, 2009.

\bibitem{sahoo2003critical}
R.~K. Sahoo, A.~J. Oliner, I.~Rish, M.~Gupta, J.~E. Moreira, S.~Ma, R.~Vilalta,
  and A.~Sivasubramaniam, ``Critical event prediction for proactive management
  in large-scale computer clusters,'' in \emph{Proceedings of the ninth ACM
  SIGKDD international conference on Knowledge discovery and data mining},
  2003, pp. 426--435.

\bibitem{yalagandula2004beyond}
P.~Yalagandula, S.~Nath, H.~Yu, P.~B. Gibbons, and S.~Seshan, ``Beyond
  availability: Towards a deeper understanding of machine failure
  characteristics in large distributed systems.'' in \emph{WORLDS}, 2004.

\bibitem{mickens2006exploiting}
J.~W. Mickens and B.~D. Noble, ``Exploiting availability prediction in
  distributed systems.'' in \emph{NSDI}, vol.~6, 2006, pp. 73--86.

\bibitem{nukada2011nvcr}
A.~Nukada, H.~Takizawa, and S.~Matsuoka, ``Nvcr: A transparent
  checkpoint-restart library for nvidia cuda,'' in \emph{2011 IEEE
  International Symposium on Parallel and Distributed Processing Workshops and
  Phd Forum}.\hskip 1em plus 0.5em minus 0.4em\relax IEEE, 2011, pp. 104--113.

\bibitem{rezaei2014snapify}
A.~Rezaei, G.~Coviello, C.-H. Li, S.~Chakradhar, and F.~Mueller, ``Snapify:
  Capturing snapshots of offload applications on xeon phi manycore
  processors,'' in \emph{Proceedings of the 23rd international symposium on
  High-performance parallel and distributed computing}, 2014, pp. 1--12.

\bibitem{el2013reading}
N.~El-Sayed and B.~Schroeder, ``Reading between the lines of failure logs:
  Understanding how hpc systems fail,'' in \emph{2013 43rd annual IEEE/IFIP
  international conference on dependable systems and networks (DSN)}.\hskip 1em
  plus 0.5em minus 0.4em\relax IEEE, 2013, pp. 1--12.

\bibitem{ghiasvand2016lessons}
S.~Ghiasvand, F.~M. Ciorba, R.~Tsch{\"u}ter, and W.~E. Nagel, ``Lessons learned
  from spatial and temporal correlation of node failures in high performance
  computers,'' in \emph{2016 24th Euromicro International Conference on
  Parallel, Distributed, and Network-Based Processing (PDP)}.\hskip 1em plus
  0.5em minus 0.4em\relax IEEE, 2016, pp. 377--381.

\bibitem{kimura2018proactive}
T.~Kimura, A.~Watanabe, T.~Toyono, and K.~Ishibashi, ``Proactive failure
  detection learning generation patterns of large-scale network logs,''
  \emph{IEICE Transactions on Communications}, 2018.

\bibitem{ghiasvand2019anomaly}
S.~Ghiasvand and F.~M. Ciorba, ``Anomaly detection in high performance
  computers: A vicinity perspective,'' in \emph{2019 18th International
  Symposium on Parallel and Distributed Computing (ISPDC)}.\hskip 1em plus
  0.5em minus 0.4em\relax IEEE, 2019, pp. 112--120.

\bibitem{fadishei2009job}
H.~Fadishei, H.~Saadatfar, and H.~Deldari, ``Job failure prediction in grid
  environment based on workload characteristics,'' in \emph{2009 14th
  International CSI Computer Conference}.\hskip 1em plus 0.5em minus
  0.4em\relax IEEE, 2009, pp. 329--334.

\bibitem{chen2014failure}
X.~Chen, C.-D. Lu, and K.~Pattabiraman, ``Failure prediction of jobs in compute
  clouds: A google cluster case study,'' in \emph{2014 IEEE International
  Symposium on Software Reliability Engineering Workshops}.\hskip 1em plus
  0.5em minus 0.4em\relax IEEE, 2014, pp. 341--346.

\bibitem{islam2017predicting}
T.~Islam and D.~Manivannan, ``Predicting application failure in cloud: A
  machine learning approach,'' in \emph{2017 IEEE International Conference on
  Cognitive Computing (ICCC)}.\hskip 1em plus 0.5em minus 0.4em\relax IEEE,
  2017, pp. 24--31.

\bibitem{andresen2018machine}
D.~Andresen, W.~Hsu, H.~Yang, and A.~Okanlawon, ``Machine learning for
  predictive analytics of compute cluster jobs,'' \emph{arXiv preprint
  arXiv:1806.01116}, 2018.

\bibitem{li2020monster}
J.~Li, G.~Ali, N.~Nguyen, J.~Hass, A.~Sill, T.~Dang, and Y.~Chen, ``Monster: An
  out-of-the-box monitoring tool for high performance computing systems,'' in
  \emph{2020 IEEE International Conference on Cluster Computing
  (CLUSTER)}.\hskip 1em plus 0.5em minus 0.4em\relax IEEE, 2020, pp. 119--129.

\bibitem{hpcc}
\BIBentryALTinterwordspacing
HPCC. (2021) {High Performance Computing Center}. [Online]. Available:
  \url{http:www.depts.ttu.edu/hpcc/}
\BIBentrySTDinterwordspacing

\bibitem{idrac}
\BIBentryALTinterwordspacing
D.~Technologies. (2021) {Integrated Dell Remote Access Controller (iDRAC)}.
  [Online]. Available:
  \url{https://www.delltechnologies.com/en-us/solutions/openmanage/idrac.htm}
\BIBentrySTDinterwordspacing

\bibitem{redfish}
\BIBentryALTinterwordspacing
DMTF. (2021) {DMTF’s Redfish}®. [Online]. Available:
  \url{https://www.dmtf.org/standards/redfish}
\BIBentrySTDinterwordspacing

\bibitem{uge}
\BIBentryALTinterwordspacing
U.~G. Engine. (2020) {Univa Grid Engine}. [Online]. Available:
  \url{https://www.univa.com/}
\BIBentrySTDinterwordspacing

\bibitem{li2006job}
H.~Li, D.~Groep, L.~Wolters, and J.~Templon, ``Job failure analysis and its
  implications in a large-scale production grid,'' in \emph{2006 Second IEEE
  International Conference on e-Science and Grid Computing
  (e-Science'06)}.\hskip 1em plus 0.5em minus 0.4em\relax IEEE, 2006, pp.
  27--27.

\bibitem{ugeaccounting}
\BIBentryALTinterwordspacing
G.~Engine. (2010) {Grid engine Man Pages}. [Online]. Available:
  \url{http://gridscheduler.sourceforge.net/htmlman/htmlman5/accounting.html}
\BIBentrySTDinterwordspacing

\bibitem{schroeder2009large}
B.~Schroeder and G.~A. Gibson, ``A large-scale study of failures in
  high-performance computing systems,'' \emph{IEEE transactions on Dependable
  and Secure Computing}, vol.~7, no.~4, pp. 337--350, 2009.

\bibitem{dummy}
\BIBentryALTinterwordspacing
Wikipedia. (2021) {Dummy variable (statistics)}. [Online]. Available:
  \url{https://en.wikipedia.org/wiki/Dummy\_variable\_(statistics)}
\BIBentrySTDinterwordspacing

\bibitem{pedregosa2011scikit}
F.~Pedregosa, G.~Varoquaux, A.~Gramfort, V.~Michel, B.~Thirion, O.~Grisel,
  M.~Blondel, P.~Prettenhofer, R.~Weiss, V.~Dubourg \emph{et~al.},
  ``Scikit-learn: Machine learning in python,'' \emph{the Journal of machine
  Learning research}, vol.~12, pp. 2825--2830, 2011.

\bibitem{zheng2011co}
Z.~Zheng, L.~Yu, W.~Tang, Z.~Lan, R.~Gupta, N.~Desai, S.~Coghlan, and
  D.~Buettner, ``Co-analysis of ras log and job log on blue gene/p,'' in
  \emph{2011 IEEE International Parallel \& Distributed Processing
  Symposium}.\hskip 1em plus 0.5em minus 0.4em\relax IEEE, 2011, pp. 840--851.

\bibitem{di2014lessons}
C.~Di~Martino, Z.~Kalbarczyk, R.~K. Iyer, F.~Baccanico, J.~Fullop, and
  W.~Kramer, ``Lessons learned from the analysis of system failures at
  petascale: The case of blue waters,'' in \emph{2014 44th Annual IEEE/IFIP
  International Conference on Dependable Systems and Networks}.\hskip 1em plus
  0.5em minus 0.4em\relax IEEE, 2014, pp. 610--621.

\bibitem{hwang2012cosmic}
A.~A. Hwang, I.~A. Stefanovici, and B.~Schroeder, ``Cosmic rays don't strike
  twice: understanding the nature of dram errors and the implications for
  system design,'' \emph{ACM SIGPLAN Notices}, vol.~47, no.~4, pp. 111--122,
  2012.

\bibitem{sridharan2013feng}
V.~Sridharan, J.~Stearley, N.~DeBardeleben, S.~Blanchard, and S.~Gurumurthi,
  ``Feng shui of supercomputer memory positional effects in dram and sram
  faults,'' in \emph{SC'13: Proceedings of the International Conference on High
  Performance Computing, Networking, Storage and Analysis}.\hskip 1em plus
  0.5em minus 0.4em\relax IEEE, 2013, pp. 1--11.

\bibitem{nie2017characterizing}
B.~Nie, J.~Xue, S.~Gupta, C.~Engelmann, E.~Smirni, and D.~Tiwari,
  ``Characterizing temperature, power, and soft-error behaviors in data center
  systems: Insights, challenges, and opportunities,'' in \emph{2017 IEEE 25th
  International Symposium on Modeling, Analysis, and Simulation of Computer and
  Telecommunication Systems (MASCOTS)}.\hskip 1em plus 0.5em minus 0.4em\relax
  IEEE, 2017, pp. 22--31.

\bibitem{fu2012hybrid}
S.~Fu, J.~Liu, and H.~Pannu, ``A hybrid anomaly detection framework in cloud
  computing using one-class and two-class support vector machines,'' in
  \emph{International conference on advanced data mining and
  applications}.\hskip 1em plus 0.5em minus 0.4em\relax Springer, 2012, pp.
  726--738.

\bibitem{nie2018machine}
B.~Nie, J.~Xue, S.~Gupta, T.~Patel, C.~Engelmann, E.~Smirni, and D.~Tiwari,
  ``Machine learning models for gpu error prediction in a large scale hpc
  system,'' in \emph{2018 48th Annual IEEE/IFIP International Conference on
  Dependable Systems and Networks (DSN)}.\hskip 1em plus 0.5em minus
  0.4em\relax IEEE, 2018, pp. 95--106.

\end{thebibliography}

\end{document}